# Origin of the Extremely Large Magnetoresistance in the Semimetal YSb

J. Xu[1,2], N. J. Ghimire[1], J. S. Jiang[1], Z. L. Xiao*[1,2], A. S. Botana[1], Y. L. Wang[1,3], Y. Hao[1,4], J. E. Pearson[1], and W. K. Kwok[1]

[1]Materials Science Division, Argonne National Laboratory, Argonne, Illinois 60439, USA

[2]Department of Physics, Northern Illinois University, DeKalb, Illinois 60115, USA

[3]Department of Physics, University of Notre Dame, Notre Dame, Indiana 46556, USA

[4]Departments of Physics, University of Illinois at Chicago, Chicago, Illinois 60607, USA

Electron-hole (*e-h*) compensation is a hallmark of multi-band semimetals with extremely large magnetoresistance (XMR) and has been considered to be the basis for XMR. Recent spectroscopic experiments, however, reveal that YSb with non-saturating magnetoresistance is uncompensated, questioning the *e-h* compensation scenario for XMR. Here we demonstrate with magnetoresistivity and angle dependent Shubnikov – de Haas (SdH) quantum oscillation measurements that YSb does have nearly perfect *e-h* compensation, with a density ratio of ~0.95 for electrons and holes. The density and mobility anisotropy of the charge carriers revealed in the SdH experiments allow us to quantitatively describe the magnetoresistance with an anisotropic multi-band model that includes contributions from all Fermi pockets. We elucidate the role of compensated multi-bands in the occurrence of XMR by demonstrating the evolution of calculated magnetoresistances for a single band and for various combinations of electron and hole Fermi pockets.

The discovery of non-saturating and extremely large magnetoresistance (XMR) in the semimetal WTe$_2$ in 2014 [1] triggered extensive research to uncover the origin of XMR, which has also been observed in many nonmagnetic materials such as bismuth [2,3], graphite [3], doped InSb [4], PtSn$_4$ [5], PdCoO$_2$ [6], NbSb$_2$ [7], and also in topological semimetals [8-18]. Besides exotic mechanisms such as topological protection [8,19,20] and magnetic-field induced metal-insulator transition (MIT) [19-23], XMR is also considered to originate from electron-hole (*e-h*) compensation [11, 24-28], as suggested by Ali et al. [1]. An isotropic semiclassical two-band model [1] with perfect *e-h* compensation (i.e., $n_e = n_h$) leads to $MR = \mu_e \mu_h H^2$, where $MR = (\rho - \rho_0)/\rho_0$ and $\rho$ and $\rho_0$ are the resistivities in the presence and absence of a magnetic field *H,* respectively, and $n_e, \mu_e$ and $n_h, \mu_h$ are the densities and mobilites of the electrons and holes, respectively. The quadratic magnetic-field dependence provides a straightforward explanation of the non-saturation behavior of the magnetoresistance, with large values of $\mu_e$ and $\mu_h$, ensuring XMR. Since a perfect (or nearly perfect) *e-h* compensation is often found in multi-band semimetals with XMR [28-31], it has become the prevalent explanation for the observed non-saturating XMR [11,24-31].

Lately, XMR was observed in the rare-earth monopnictides LnX (Ln = La/Y/Nd/Ce and X = Sb/Bi) [19,24-28,30-36]. Similar to other multi-band XMR materials, a nearly perfect *e-h* compensation was found in LaSb and LaBi and proposed as one of the origins for their XMR behavior [24-28,30]. However, recent angle-resolved photoemission spectroscopy (ARPES) measurements on YSb crystals [32], revealed a ratio of 0.81 for the electron/hole concentration. In the semiclassical two-band model, such an imbalance in charge carrier densities will result in a saturating MR behavior at intermediate magnetic fields, unless a large difference in the *e-h* mobility ($\mu_e/\mu_h > 250$) is invoked to account for the observed XMR [32]. Since YSb is topologically trivial [32] and the MIT mechanism can also be excluded [33-34], the ARPES finding indicated a new origin for the observed XMR.

Here we report on magnetoresistivity measurements to uncover the origin of the non-saturating XMR in YSb. We measured angle dependent Shubnikov – de Haas (SdH) oscillation to determine the shape and volume of the Fermi surface. Like in ARPES, our measurement shows anisotropic electron pockets and nearly isotropic hole pockets. However, in contrast to the ARPES result, our quantum oscillations data suggests a nearly compensated *e-h* ratio ($n_e/n_h \approx 0.95$) in YSb. More importantly, we show that a semiclassical theory can quantitatively describe both the transverse MR and Hall resistance when contributions from both the anisotropic (electron) and isotropic (hole) Fermi pockets are included.

We measured two YSb crystals grown in Sb self flux, with more synthesis and characterization details given in Ref.25. The dimensions of the crystals are 473.6 μm (*w*) × 85 μm (*d*) × 493.5 μm (*l*) and 950 μm (*w*) × 180 μm (*d*) × 982.5 μm (*l*) for samples I and II, respectively, with *d* and *w* being the thickness and width of the crystal and *l* the separation between the two voltage contacts for transverse magnetoresistivity measurements. DC resistivity experiments [37,38] were conducted in Quantum Design PPMS (PPMS-9) using constant current mode. Angular dependent measurements were realized by placing the sample on a precision, stepper-controlled rotator with an angular resolution of 0.05°, with the magnetic field $H(\theta)$ rotated in the *y-z*, i.e. (100) plane and the applied current *I* along the *x*- ([100]) direction, such that the magnetic field is always perpendicular to the current as indicated in the inset of Fig.1.

In Fig.1 and Fig.S1 [37] we present the temperature dependence of the zero-field resistivity $\rho_0(T)$ and $\rho_{xx}(H)$ at *T* = 5 K, respectively, for sample I. We obtained a residual resistivity ratio *rrr* of ~200 and a *MR* of $1.3 \times 10^5$ %, indicating the high quality of the crystal. The *MR* in our YSb crystals follows a power-law dependence with magnetic field, with an exponent less than but close to 2 (see inset of Fig.S1 [37]), similar to other XMR materials [38-41].

The bulk electronic band structure and Fermi surface of rare-earth monopnictides LnX were investigated more than three decades ago [42-44] and also reported in recent publications [19,24-26,30,32,33]. The bulk FS consists of electron pockets centered at $X$ and elongated along the $\Gamma-X$ direction in addition to two hole pockets centered at $\Gamma$ [24–26]. As revealed by ARPES [32] and illustrated by the projection of the calculated Fermi pockets on the field-rotation (100) plane in the inset of Fig.2(b), the electronic structure of YSb has three ($\alpha, \beta$ and $\gamma$) bands [25]. The electron band, $\alpha$, has three orthogonally arranged ellipsoidal Fermi pockets ($\alpha_1, \alpha_2,$ and $\alpha_3$) while the hole bands $\beta$ and $\gamma$ Fermi pockets are nearly spherical. We conducted angle-dependent SdH oscillation measurements to determine these Fermi pockets and to obtain the charge carrier density and the mobility anisotropy.

Fig.2(a) shows a typical $\rho_{xx}(H)$ curve at $T$ = 2.5 K and $\theta$ =139.5°. SdH oscillations can be observed at high fields, and highlighted in the inset after subtracting a smooth background. FFT analysis result is presented in Fig.2(b). We identified four fundamental frequencies and their higher harmonics that can be indexed to the $\alpha_1, \alpha_2, \beta$ and $\gamma$ Fermi pockets shown in the inset. We did not observe frequencies expected from the $\alpha_3$ Fermi pocket. This absence, however, is not difficult to understand: the current flows along the long axis of the elliptical $\alpha_3$ Fermi pocket, and hence the mobility of the associated electrons is low [see discussions below: the mobility ($\mu_\parallel$) of the electrons from the $\alpha_3$ Fermi pocket is a factor of ~10 (=$\lambda_\mu^2$) smaller than that ($\mu_\perp$) of the $\alpha_1$ and $\alpha_2$ Fermi pockets]. Since the oscillation amplitude depends exponentially on the mobility [24], $\Delta\rho \sim e^{-1/\mu H}$, the SdH quantum oscillations from the $\alpha_3$ Fermi pocket could be below the measurement sensitivity level associated with our maximum magnetic field of 9 Tesla.

Applying the same analysis procedure to the SdH oscillations obtained at various angles, we derive the angle dependences of the frequency $F$ for the three bands, as shown in Fig.3. Quantitatively, the angle dependence of $F_{\alpha 1}, F_{\alpha 2}$ can be fitted with

$$F_\alpha = F_0/\sqrt{\cos^2[\theta - (n-1)\pi/2] + \lambda_\mu^{-2}\sin^2[\theta - (n-1)\pi/2]}$$

where $F_0 = 355$ Tesla, $\lambda_\mu = 3.26$, and $n = 1, 2$ for the $\alpha_1$, $\alpha_2$ pockets, respectively.

Using the Onsager relation $F = (\phi_0/2\pi^2)A$ with $\phi_0$ and $A$ being the flux quantum and the area of the extremal orbit [24], we can extract the short Fermi vector $k_F^S = 1.0386 \times 10^7$ cm$^{-1}$ and the long Fermi vector $k_F^L = \lambda_\mu k_F^S = 3.3858 \times 10^7$ cm$^{-1}$ for the electron ellipsoid pocket. This leads to a density of $n_e^\alpha = 1.2335 \times 10^{20}$ cm$^{-3}$ for $\alpha_1$ and $\alpha_2$. Due to the crystal symmetry, however, the three electron pockets are equivalent, and hence $\alpha_3$ should have the same electron density as $\alpha_1$ and $\alpha_2$, resulting in a total electron density of $n_e = 3.7006 \times 10^{20}$ cm$^{-3}$.

The frequencies for the two hole pockets show a slight angle dependence with a four-fold symmetry. Mathematically, we can fit the data for the $\beta$ and $\gamma$ pockets respectively with

$$F_\beta = 728/\sqrt{\cos^2(2\theta - \pi/2) + 1.015^{-2}\sin^2(2\theta - \pi/2)} \text{ and}$$

$$F_\gamma = 1277/\sqrt{\cos^2(2\theta - \pi/2) + 1.115^{-2}\sin^2(2\theta - \pi/2)}$$

The 'anisotropy' of 1.015 and 1.115 for the $\beta$ and $\gamma$ pockets, respectively, is much smaller than that (3.26) of the $\alpha$ pockets. To calculate the hole density, we treat both $\beta$ and $\gamma$ Fermi pockets as spheres with average values $F_\beta = 732$ T and $F_\gamma = 1351$ T of the minimal and maximal frequencies for the $\beta$ and $\gamma$ pockets, corresponding to hole densities of $n_h^\beta = 1.1204 \times 10^{20}$ cm$^{-3}$ and $n_h^\gamma = 2.8091 \times 10^{20}$ cm$^{-3}$, respectively. This results in a total hole density of $n_h = 3.9295 \times 10^{20}$ cm$^{-3}$. Thus, we obtain an electron-hole ratio of $n_e/n_h = 0.942$ for sample I. The SdH results for sample II are presented in Fig.S2, yielding $n_e/n_h = 0.949$. That is, both samples give consistent results and reveal that YSb is a compensated semimetal.

As discussed in the introduction and also in Ref.45, $\rho_{xx}(H)$ and $\rho_{xy}(H)$ have typically been described using an isotropic two-band model, which assumes the same mobility in all directions for each type of charge carriers, i.e., $\mu_e$ for all electrons and $\mu_h$ for all holes. Clearly, the assumption on electron mobility is not applicable to YSb, where the mobility of electrons from the three anisotropic Fermi pockets differs significantly. For comparison purpose, we present in Fig.4(a) the results for our YSb crystal using the two-band model. We obtain $n_e/n_h = 0.982$, $\mu_e = 0.935$ m²V⁻¹s⁻¹ and $\mu_h = 1.056$ m²V⁻¹s⁻¹, which are comparable to those of LaSb [24]. The derived charge carrier densities ($n_e = 1.202 \times 10^{20}$ cm$^{-3}$, $n_h = 1.223 \times 10^{20}$ cm$^{-3}$), however, are only 1/3 of those obtained from SdH measurements.

For a magnetic field $H$ applied in the $z$-direction with current flow along the $x$-axis, the magnetoconductivity tensor for an anisotropic electron Fermi pocket is given as follows [46]:

$$\hat{\sigma} = \begin{pmatrix} \sigma_{xx} & \sigma_{yx} \\ \sigma_{xy} & \sigma_{yy} \end{pmatrix} \quad (1)$$

with $\sigma_{xx} = ne\mu_x/(1 + \mu_x\mu_y H^2)$; $\sigma_{yy} = ne\mu_y/(1 + \mu_x\mu_y H^2)$; $\sigma_{yx} = -\sigma_{xy} = ne\mu_x\mu_y H/(1 + \mu_x\mu_y H^2)$. Here, $n$ is the electron density, and $\mu_x$ and $\mu_y$ are the respective mobilities along the $x$ and $y$ axes. Eq.(1) is applicable for an ellipsoidal hole pocket by changing the sign of both the charge $e$ and the mobility. It can also be implemented for the isotropic case by assuming $\mu_x = \mu_y$. By replacing $\sigma_{ij}$ in Eq.(1) with $\sigma_{ij}^T = \sum_p \sigma_{ij}^p$ where $p = \alpha_1, \alpha_2, \alpha_3, \beta,$ and $\gamma$, we obtain the magnetoresistivity tensor [46]:

$$\hat{\rho} = \begin{pmatrix} \rho_{xx} & \rho_{yx} \\ \rho_{xy} & \rho_{yy} \end{pmatrix} \quad (2)$$

where $\rho_{xx} = \sigma_{yy}^T/[\sigma_{xx}^T\sigma_{yy}^T + (\sigma_{xy}^T)^2]$, $\rho_{yy} = \sigma_{xx}^T/[\sigma_{xx}^T\sigma_{yy}^T + (\sigma_{xy}^T)^2]$, $\rho_{yx} = -\rho_{xy} = \sigma_{xy}^T/[\sigma_{xx}^T\sigma_{yy}^T + (\sigma_{xy}^T)^2]$.

Since $\sigma_{ij}^T$ includes contributions from all five Fermi pockets, Eq.(2) provides a complete description of the measured transverse and Hall magnetoresistivities $\rho_{xx}(H)$ and $\rho_{xy}(H)$. From Eq.(1) we know that the magnetoconductivity of each Fermi pocket is determined by three parameters ($n, \mu_x$ and $\mu_y$). Once the ratio $k_F^L/k_F^S$ of the ellipse's semimajor and semiminor axes $k_F^L$ and $k_F^S$ is known, the relationship of the mobility along the long axis $\mu_\parallel$ and the short axis $\mu_\perp$ can be described as $\mu_\perp/\mu_\parallel = m_\parallel/m_\perp = (k_F^L/k_F^S)^2$, where $m_\parallel$ and $m_\perp$ are the effective masses along the long and short axes [47]. That is, only one of the two mobilities is an independent fitting parameter. Due to the crystalline symmetry, $\alpha_1$, $\alpha_2$ and $\alpha_3$ have identical Fermi pockets but oriented differently. Their electron densities $n_1$, $n_2$ and $n_3$ are the same, i.e., $n_1 = n_2 = n_3 = n_e^\alpha$, and equal to one third of the total electron density $n_e$. For example, we can rewrite the magnetoconductivity for the $\alpha_1$, $\alpha_2$ and $\alpha_3$ Fermi pocket $\sigma_{xx}^{\alpha 1} = n_e^\alpha e\mu_\perp/(1 + \mu_\perp^2 H^2)$, $\sigma_{xy}^{\alpha 1} = n_e^\alpha e\mu_\perp^2 H/(1 + \mu_\perp^2 H^2)$; $\sigma_{xx}^{\alpha 2} = n_e^\alpha e\mu_\perp/(1 + \mu_\perp^2 H^2/\lambda_\mu^2)$, $\sigma_{xy}^{\alpha 2} = n_e^\alpha e\mu_\perp^2 H/(\lambda_\mu^2 + \mu_\perp^2 H^2)$; and $\sigma_{xx}^{\alpha 3} = n_e^\alpha e\mu_\perp/(\lambda_\mu^2 + \mu_\perp^2 H^2)$, $\sigma_{xy}^{\alpha 3} = n_e^\alpha e\mu_\perp^2 H/(\lambda_\mu^2 + \mu_\perp^2 H^2)$, with $\lambda_\mu = k_F^L/k_F^S$. That is, we have only two independent fitting parameters ($n_e^\alpha, \mu_\perp$) for the three electron Fermi pockets. For simplification we treat the hole Fermi pockets as a sphere with an isotropic mobility of $\mu_x = \mu_y = \mu_h$ and a zero-field conductivity of $\sigma_{xx}^{h0} = n_h e\mu_h$. With these two parameters we can obtain $\sigma_{ij}$ for the hole bands, e.g., $\sigma_{xx}^\beta = n_h^\beta e\mu_h^\beta/[1 + (\mu_h^\beta H)^2]$ and $\sigma_{xy}^\beta = n_h^\beta e(\mu_h^\beta)^2 H/[1 + (\mu_h^\beta H)^2]$. Thus, we have six free variables $n_e^\alpha, \mu_\perp$; $n_h^\beta, \mu_h^\beta$ and $n_h^\gamma, \mu_h^\gamma$ for the five Fermi pockets.

We found that the fit of Eq.(2) to the experimental data is very sensitive to the value of the carrier density. We were unable to achieve reasonable fits to both $\rho_{xx}(H)$ and $\rho_{xy}(H)$ by applying the experimental $n_e^\alpha$, $n_h^\beta$ and $n_h^\gamma$. Since first principles calculations give a more complicated $\gamma$ Fermi pocket than the one we determined from its projection on the (100) plane, our estimated density $n_h^\gamma$ could have significant deviation. In the analysis we treat $n_h^\gamma$ as a free variable.

Using the experimentally determined $n_e^\alpha$, $n_h^\beta$, we can use Eq.(2) to quantitatively describe both $\rho_{xx}(H)$ and $\rho_{xy}(H)$, as shown in Fig.4(b). It gives $n_h^\gamma = 2.558 \times 10^{20} \text{cm}^{-3}$, $\mu_\perp = 11.96 \text{ m}^2\text{V}^{-1}\text{s}^{-1}$, $\mu_h^\beta = 9.64 \text{ m}^2\text{V}^{-1}\text{s}^{-1}$, $\mu_h^\gamma = 2.482 \text{ m}^2\text{V}^{-1}\text{s}^{-1}$. These mobility values differ significantly from the $\mu_e$ and $\mu_h$ derived using the isotropic two-band model, demonstrating the necessity to include the anisotropy and multi-band nature of the material in a quantitative MR analysis. For comparison, we present in Table S1 [37] a summary of the density and mobility of the charge carriers derived from the SdH measurements, the isotropic two-band model and anisotropic multi-band model. We note that the derived $n_h^\gamma = 2.558 \times 10^{20} \text{cm}^{-3}$ obtained through analysis with Eq.(2) is ~ 9% smaller than that ($2.8091 \times 10^{20}$ cm$^{-3}$) from SdH measurements, resulting in $n_e/n_h$ = 1.006. This indicates that YSb may indeed be a perfect compensated system.

To further demonstrate the role played by the compensated multi-bands on XMR, we show in Fig.5 the calculated $\rho_{xx}$ and $\rho_{xy}$ for a single Fermi pocket and combinations thereof using the parameters obtained from above analysis using Eq.(2). If only the Fermi pocket $\alpha_1$ were present, we find a field independent $\rho_{xx}$ and hence the absence of a MR. However, the addition of pockets $\alpha_2$ and $\alpha_3$ results in a magnetic field-dependent $\rho_{xx}$ with a $MR \approx 300\%$ at $\mu_0 H$ = 9 T. $\rho_{xx}$ is further enhanced by adding the hole pockets, resulting in $\rho_{xx}$ to reach an extremely large value of $1.3\times10^5$ %.

In summary, SdH quantum oscillation measurement shows a compensated electron-hole behavior in YSb. We find that the origin of the non-saturating XMR in YSb is still semiclassical without requiring the presence of surface conduction. Both the transverse and Hall magnetoresistivies can be described with an anisotropic multi-band model that allows contributions from all electron and hole Fermi pockets. We demonstrated the importance of the coexistence of multiple electron and hole bands in the occurrence of XMR.

**ACKNOWLEDGEMENTS**

This work was supported by the U.S. DoE, Office of Science, Basic Energy Sciences, Materials Sciences and Engineering Division. Partial magnetoresistance measurements were also carried out at Northern Illinois University under NSF Grant No. DMR-1407175.  The micro-contacting was performed at Argonne's Center for Nanoscale Materials (CNM) which is supported by DOE BES under Contract No. DE-AC02-06CH11357. We thank John Mitchell and Michael R. Norman for stimulating discussions.

*Corresponding author, xiao@anl.gov or zxiao@niu.edu

**Figure captions**

**FIG.1.** (color online) Temperature dependence of the resistivity at zero field. The lower right panel presents the same data in the semi-log plot to show the large residual resistivity ratio *rrr* (~200). The upper left inset is a schematic defining the angle $\theta$ of the magnetic field orientation.

**FIG.2.** (color online) (a) $\rho_{xx}(H)$ curve at $T = 2.5$K and $\theta = 139.5°$. The inset gives the Shubnikov-de Haas (SdH) quantum oscillation data after subtracting a smooth background. (b) Fast Fourier transform (FFT) analysis of the SdH results. Inset gives the projection of the calculated Fermi pockets in the magnetic field rotation plane – the (100) plane.

**FIG.3.** Angle dependence of the SdH oscillation frequencies. Symbols are experimental data and lines are fits to the equations described in the text for the angle dependences of $F_\alpha, F_\beta$ and $F_\gamma$ (solid and dashed lines are for the fundamental frequencies and higher harmonics, respectively).

**FIG.4.** (color online) Analysis of the Hall and longitudinal magnetoresistivities for sample I at $T = 5$ K and $H$ //c: (a) with the isotropic two-band model using $n_e = 1.202 \times 10^{20}$ cm$^{-3}$, $n_h = 1.223 \times 10^{20}$ cm$^{-3}$, $\mu_e = 0.935$ m$^2$V$^{-1}$s$^{-1}$ and $\mu_h = 1.056$ m$^2$V$^{-1}$s$^{-1}$, (b) with the anisotropic multi-band model Eq.(2) using $n_e^\alpha$ and $n_h^\beta$ determined through SdH measurements and $n_h^\gamma = 2.558 \times 10^{20}$cm$^{-3}$, $\mu_\perp = 11.96$ m$^2$V$^{-1}$s$^{-1}$, $\mu_h^\beta = 9.64$ m$^2$V$^{-1}$s$^{-1}$ and $\mu_h^\gamma = 2.482$ m$^2$V$^{-1}$s$^{-1}$. Symbols are experimental data and solid lines are the fits.

**FIG.5.** (color online) Calculated $\rho_{xx}$ and $\rho_{xy}$ using Eqs.(1)-(3) with parameters derived from the analysis of data in Fig. 4 with Eq. (2). (a) single Fermi pocket $\alpha_1$, (b) multiple Fermi pockets $\sum_{i=1}^{3} \alpha_i$, and (c) all five electron and hole Fermi pockets. The top panel is the projection of the corresponding Fermi pockets in the orbital (001) plane.

**FIGURE 1**

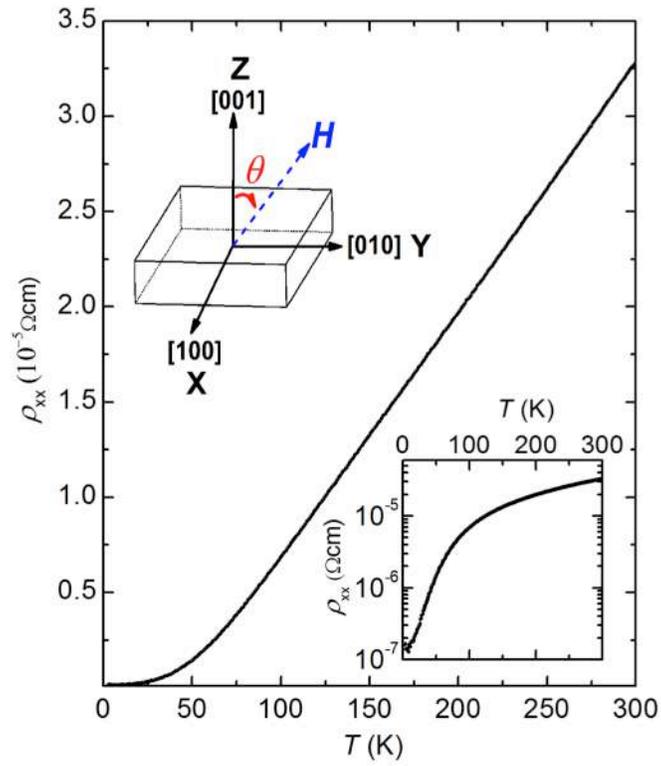

**FIGURE 2**

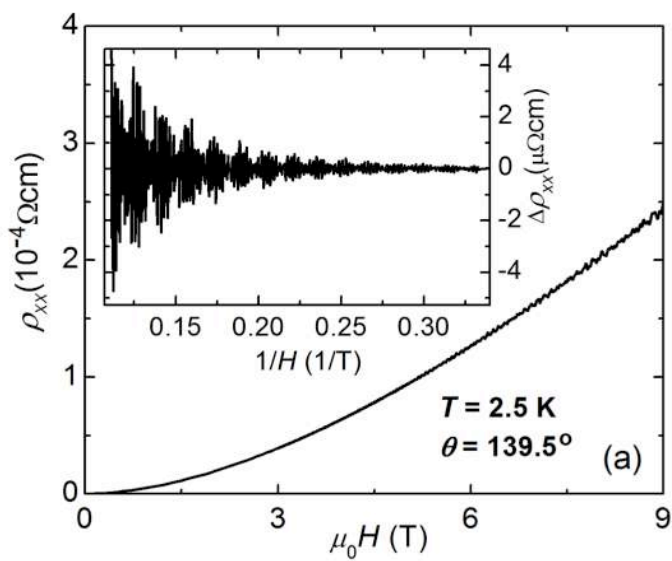

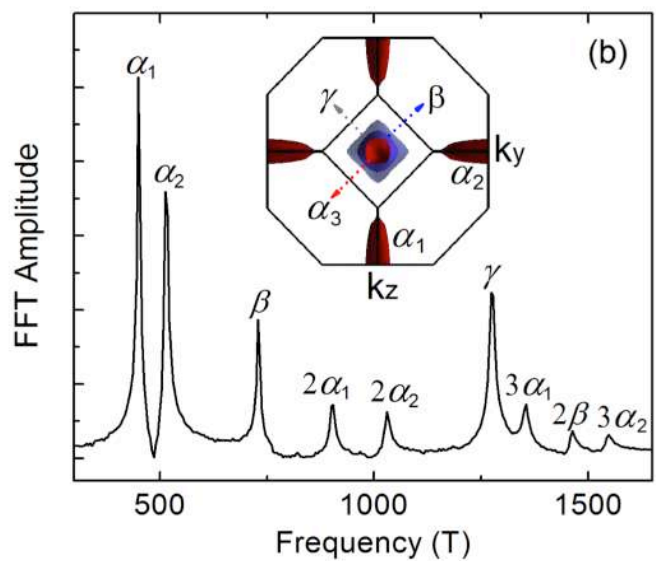

**FIGURE 3**

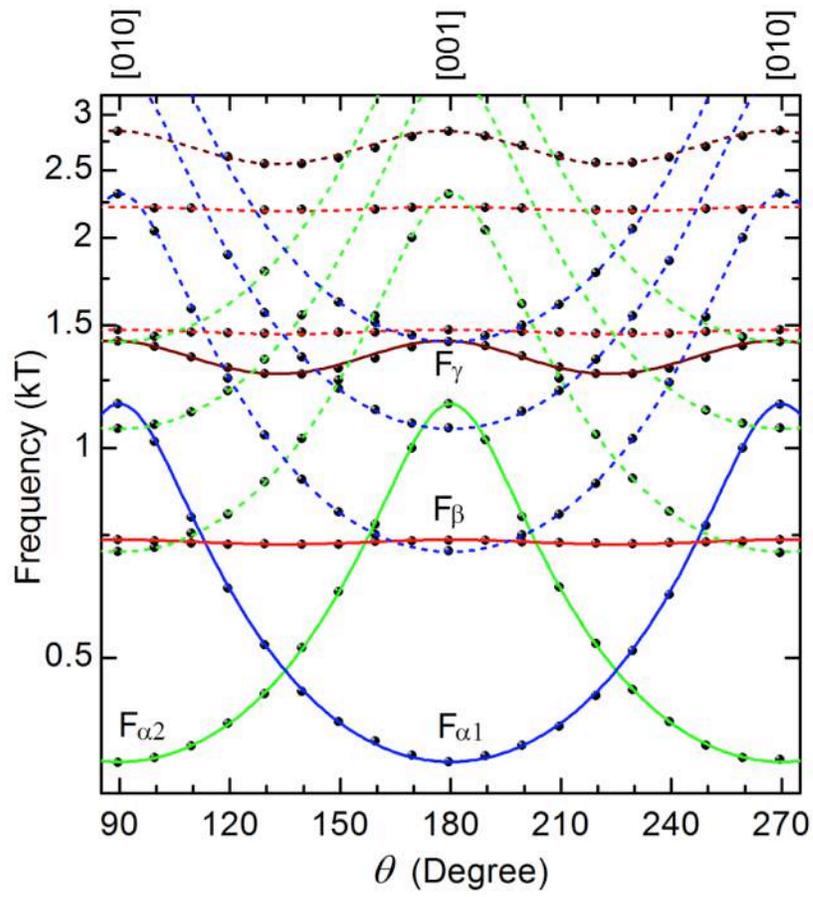

**FIGURE 4**

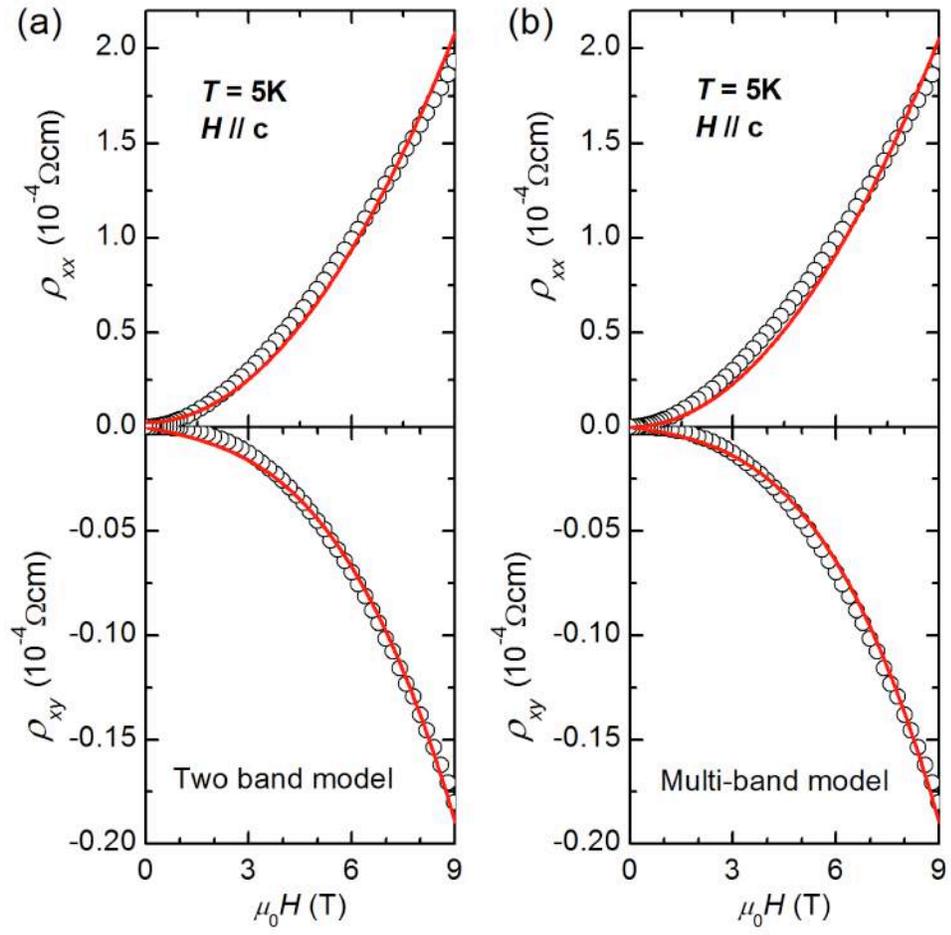

**FIGURE 5**

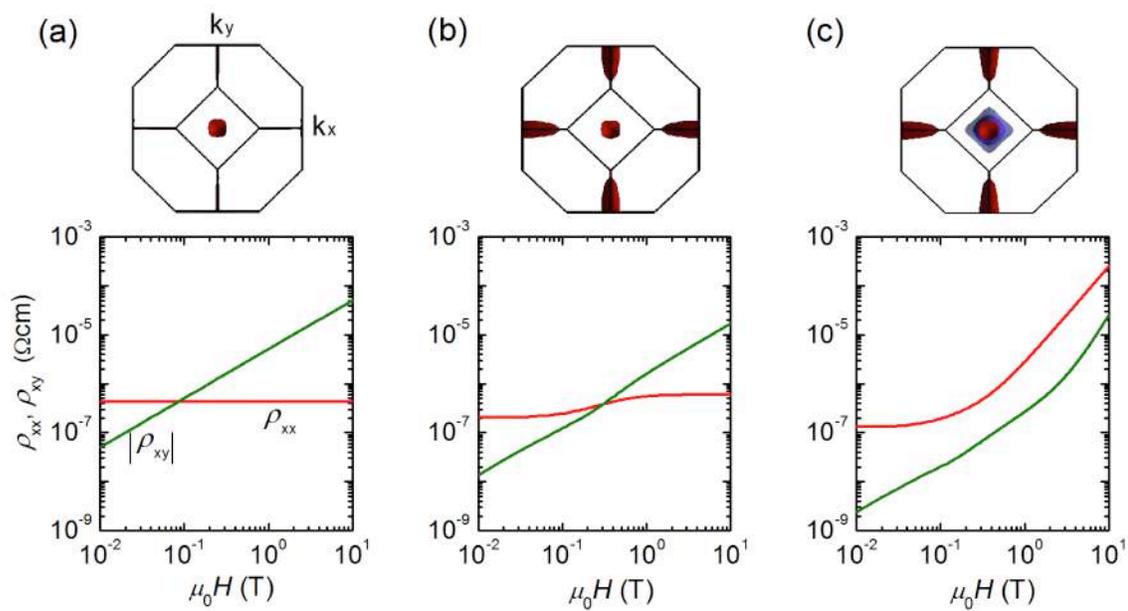

# Supplemental materials for

# Origin of the Extremely Large Magnetoresistance in the Semimetal YSb
# by J. Xu et al.

Table S1. Summary of the derived density and mobility of sample I

| | Density | Density ratio | Mobility |
|---|---|---|---|
| **SdH quantum oscillations** | $n_e = 3.7006 \times 10^{20}$ cm$^{-3}$<br>$n_h = 3.9295 \times 10^{20}$ cm$^{-3}$ | $n_e/n_h = 0.942$ | |
| **Isotropic two-band model** | $n_e = 1.202 \times 10^{20}$ cm$^{-3}$<br>$n_h = 1.223 \times 10^{20}$ cm$^{-3}$ | $n_e/n_h = 0.982$ | $\mu_e = 0.935$ m$^2$V$^{-1}$s$^{-1}$<br>$\mu_h = 1.056$ m$^2$V$^{-1}$s$^{-1}$ |
| **Anisotropic multi-band model** | $n_e = 3.7006 \times 10^{20}$ cm$^{-3}$<br>$n_h = 3.6784 \times 10^{20}$ cm$^{-3}$ | $n_e/n_h = 1.006$ | $\mu_\perp = 11.96$ m$^2$V$^{-1}$s$^{-1}$<br>$\mu_h^\beta = 9.64$ m$^2$V$^{-1}$s$^{-1}$<br>$\mu_h^\gamma = 2.482$ m$^2$V$^{-1}$s$^{-1}$ |

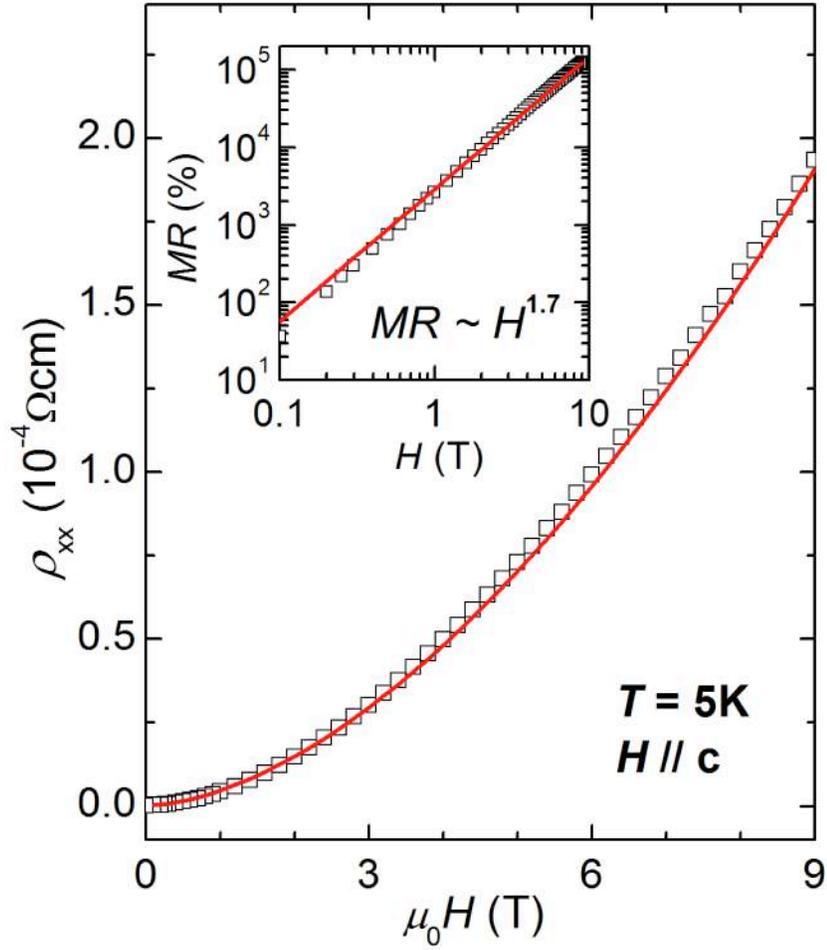

**Fig.S1.** Transverse magnetoresistivity $\rho_{xx}(H)$ at $T$ = 5K and $H$ // [001]. The solid line is a fit with $\rho_{xx}(H) = \rho_{xx}(0)[1+aH^m]$ where $m$ = 1.7 and $a$ = 29. Inset shows the corresponding *MRs*.

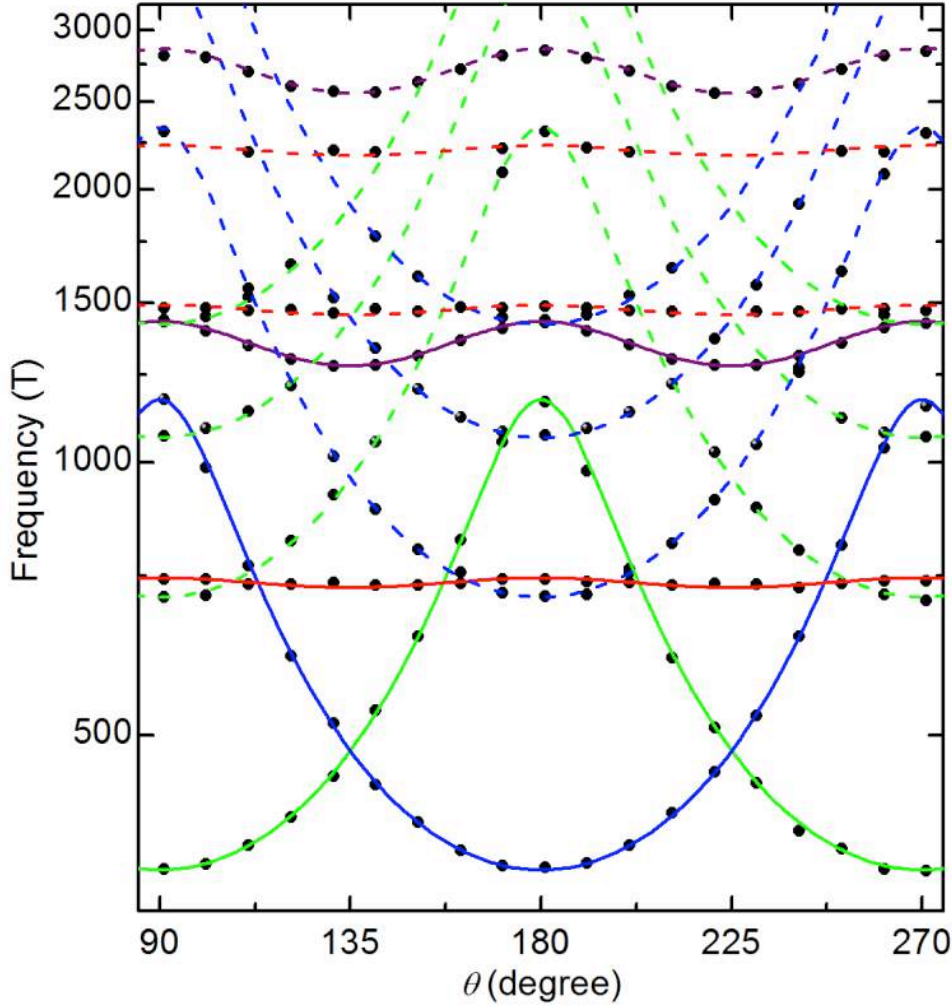

**Fig.S2**. Angle dependence of the SdH oscillation frequencies of sample II. The darker symbols and lines represent the fundamental frequencies and the lighter symbols and the dashed lines are their corresponding higher harmonics. For $F_{\alpha 1}$, $F_{\alpha 2}$ we have

$F_\alpha = F_0/\sqrt{\cos^2[\theta-(n-1)\pi/2] + \lambda_\mu^{-2}\sin^2[\theta-(n-1)\pi/2]}$, with $F_0$ = 355 Tesla, $\lambda_\mu$ = 3.3 and $n$ = 1, 2 for the $\alpha_1$ and $\alpha_2$ Fermi pockets. For $\beta$ and $\gamma$ bands we have

$F_\beta = 727/\sqrt{\cos^2 2\theta + 1.025^{-2}\sin^2 2\theta}$, $F_\gamma = 1277/\sqrt{\cos^2(2\theta-\pi/2) + 1.12^{-2}\sin^2(2\theta-\pi/2)}$.

$n_e^\alpha = 1.24866\times 10^{20}$ cm$^{-3}$, i.e., $n_e = 3.74597\times 10^{20}$ cm$^{-3}$; $n_h^\beta = 1.12954\times 10^{20}$ cm$^{-3}$, $n_h^\gamma = 2.8169\times 10^{20}$ cm$^{-3}$, i.e., $n_h = 3.94646\times 10^{20}$ cm$^{-3}$, resulting in a ratio of $n_e/n_h = 0.949$.